\documentclass[aps,prd,numerical,showpacs,showkeys,10pt]{revtex4-1}

\usepackage{amsmath}
\usepackage{amssymb}
\usepackage{amsfonts}
\usepackage{graphicx}
\usepackage{dcolumn}
\usepackage{bm}
\usepackage{hyperref}
\usepackage{wasysym}
\hypersetup{colorlinks=false}

\usepackage{graphics}
\usepackage{graphicx}
\usepackage{color}

\definecolor{LightCyan}{rgb}{0.88,1,1}
\definecolor{maroon}{cmyk}{0,0.87,0.68,0.32}
\definecolor{gray75}{gray}{0.75}
\definecolor{darkblue}{rgb}{0.089,0.21,0.363}
\definecolor{niceblue}{cmyk}{1,0.2,0.2,0.5}

\definecolor{verylightgray}{gray}{0.95}

\begin{document}
%\doublespacing

\title[Quasibound states of Schwarzschild acoustic black holes]{Quasibound states of Schwarzschild acoustic black holes}

\date{\today}

\author{H. S. Vieira}
\email{horacio.santana.vieira@hotmail.com and horacio.santana-vieira@tat.uni-tuebingen.de}
\affiliation{Theoretical Astrophysics, Institute for Astronomy and Astrophysics, University of T\"{u}bingen, 72076 T\"{u}bingen, Germany}
\author{Kostas D. Kokkotas}
\email{kostas.kokkotas@uni-tuebingen.de}
\affiliation{Theoretical Astrophysics, Institute for Astronomy and Astrophysics, University of T\"{u}bingen, 72076 T\"{u}bingen, Germany}

\begin{abstract}
In this paper, we study the recently proposed Schwarzschild acoustic black hole spacetime, in which we investigate some physical phenomena related to the effective geometry of this background, including the analogous Hawking radiation and the quasibound states. We calculate the spectrum of quasibound state frequencies and the wave functions on the Schwarzschild acoustic background by using the polynomial condition of the general Heun function, and then we discuss the stability of the system. We also compare the resonant frequencies of the Schwarzschild acoustic black hole with the ones by the standard Schwarzschild black hole. Our results may shed some light on the physics of black holes and their analog models in condensed matter. Moreover, these studies could provide the possibility, in principle, for laboratory testing of effects whose nature is unquestionably associated with purely quantum effects in gravity.
\end{abstract}



\preprint{Preprint submitted to Physical Review D}
%\preprint{AIP/123-QED}

\maketitle

%\begin{quotation}
%...
%\end{quotation}

%
%%%%%%%%%%%%%%%%%%%%%%%%%%%%%%%%%%%%%%%%%%%%%%%%%%%%%%%%%%%%%%%%%%%%%%%%%%%%%%%%%%%%%%%%%%%%%% Introduction
%
\section{Introduction}\label{I}
In the last 5 years, gravitational physics took a significant boost with the detections of gravitational waves from  neutron stars and black holes collisions and the measurements of the properties of the central black hole in M87 \cite{PhysRevLett.116.221101,AstrophysJLett.875.L1}. This progress was based in the advance technology developed for the specific observations. Still this tremendous technological advance does not apply to all phenomena associated to gravitational physics and especially the ones related to the interaction between quantum particles and the gravitational field generated by astrophysical objects. From the theoretical point of view, one needs a very accurate theory to compare and describe such events. This calls for a more systematic search on these objects, which will be carried out in the next decades \cite{ClassQuantumGrav.36.143001,GenRelatGravit.52.81}, while at the same time wherever possible, especially for quantum phenomena, the analog gravity can provide a useful testing tool \cite{PhysRevLett.117.271101,PhysRevLett.117.121301,NaturePhysics.13.833,ModPhysLettA.35.2050042,PhilTransRSocA.378.20190239,PhysRevLett.124.141101}.

In order to best understand and analyze some important quantum gravity phenomena in the vicinity of black holes, Unruh \cite{PhysRevLett.46.1351} proposed that several condensate matter systems, which present an effective geometry, could mimic some astrophysical scenarios. This is possible, in principle, by modeling the behavior of quantum particles in a classical gravitational field as the motion of sound waves in a convergent fluid flow. Thus, in the analog models of gravity, the sound waves are propagating on a hydrodynamic background, generated by acoustic black holes. This interaction is described for an equation of motion which is formally identical to the covariant massless Klein-Gordon equation. Along this line of research, many different physical phenomena were examined in the past on backgrounds with such an effective geometry \cite{PhysLettB.737.6,PhysRevD.91.104038,ModPhysLettA.32.1750047,ModPhysLettA.32.1750119,PhysRevD.96.064027,IntJModPhysA.33.1850185,JMathPhys.59.072502,ContemporaryMathematics.734.14766,CommMathPhys.378.467,IntJModPhysD.29.2041018,PhysRevD.102.124019,ModPhysLettA.35.2050236,ApplSci.10.24}.

Among these studies, we can mention two in particular, namely, the one where the Hawking-Unruh effect is probed \cite{NatureComm.10.3030} and the work where an analytic solution for the massless Klein-Gordon equation in the canonical acoustic black hole spacetime was obtained in terms of the general Heun functions \cite{GenRelatGravit.52.72}. In these works, some analog gravity phenomena were proposed and studied, which can, in principle, lead to the realization of accurate experiments trying to probe some features of the quantum field theory in curved spacetime. More specifically, if a physical phenomenon occurs in an astrophysical scenario, it will also occur in an analog gravity model which presents an effective geometry.

In this work, we focus on two physical phenomena that occur in analog black holes, namely, the computation of (i) the Hawking radiation and (ii) the quasibound state frequencies of massless scalars in the Schwarzschild acoustic black hole spacetime. The Hawking radiation is a thermal spectrum related to the quantum mechanical effects on the exterior event horizon of a black hole \cite{CommMathPhys.43.199}, and the quasibound states are solutions to the equation of motion which tend to zero at infinity \cite{PhysRevD.76.084001}, whose resonant frequency spectrum is related to the decay of the perturbation; that is, they correspond to damped oscillations. Here, the results depend on a single parameter, namely, the \emph{tuning parameter} $\xi$ associated with the fluid velocity. The Hawking radiation spectrum and the resonant frequencies will be obtained by imposing the appropriated boundary conditions on the radial equation. The solution of the radial equation will be given in terms of the general Heun functions, which are special functions of mathematical physics applied in various problems associated to wavelike equations on science in general (in particular, for applications in physics, see Ref.~\cite{AdvHighEnergyPhys.2018.8621573} and references therein).

Actually, it is worth mentioning some physical systems where the framework of Schwarzschild acoustic black holes has been implemented. There are two  experiments reporting  realizations of acoustic black holes; the first was in a Bose-Einstein condensate \cite{NatPhys.10.864}, and the second was reported in an optical medium \cite{PhysRevLett.122.010404}. Moreover, the oscillatory part of the quasinormal modes was detected for the first time in an analog black hole experiment by Torres \textit{et al.} \cite{PhysRevLett.125.011301}, in which they set up a vortex flow out of equilibrium to observe its quasinormal oscillations. Therefore, we hope that, in a near future, the Hawking radiation and the quasibound states of Schwarzschild acoustic black holes could be detected in some hydrodynamic or/and condensate matter systems, or even in a true astrophysical system.

This paper is organized as follows. In Sec. \ref{II}, we introduce the metric corresponding to the Schwarzschild acoustic black hole spacetime. In Sec. \ref{III}, we solve the covariant massless Klein-Gordon equation in the background under consideration. In Sec. \ref{IV}, we obtain the Hawking radiation spectrum. In Sec. \ref{V}, we compute the resonant frequency spectrum related to the quasibound states. In Sec. \ref{VI}, we provide the eigenfunctions by using some properties of the general Heun functions. Finally, in Sec. \ref{VII}, the conclusions are given. Here, we adopt the natural units where $G \equiv c \equiv \hbar \equiv 1$.
%
%%%%%%%%%%%%%%%%%%%%%%%%%%%%%%%%%%%%%%%%%%%%%%%%%%%%%%%%%%%%%%%%%%%%%%%%%%%%%%%%%%%%%%%%%%%%%% Schwarzschild acoustic black hole spacetime
%
\section{Schwarzschild acoustic black hole spacetime}\label{II}
In a recent paper, Ge \textit{et al.} \cite{PhysRevD.99.104047} obtained a class of solutions for analog gravity models by considering the relativistic Gross-Pitaevskii \cite{NuovoCimento.20.454, SovPhysJETP.13.451} and Yang-Mills \cite{PhysRev.96.191} theories. The constructed metrics correspond to acoustic black holes. Among these metrics, we are interested in a particular one, namely, the one describing the Schwarzschild acoustic black hole spacetime. In what follows, we will review the properties of this acoustic black hole solution, on which we want to investigate the behavior of scalar fields following Ref.~\cite{PhysRevD.99.104047}.

The action of the Gross-Pitaevskii theory describing a nonlinear scalar field is given by
\begin{equation}
S=\int d^{4}x\ \sqrt{-g}\ \biggl(|\partial_{\mu}\varphi|^{2}+m^{2}|\varphi|^{2}-\frac{b}{2}|\varphi|^{4}\biggr),
\label{eq:action_GP}
\end{equation}
where $\varphi$ is a complex scalar field as order parameter; $m^{2}(\sim T-T_{c})$ is a temperature(-dependent) parameter, with $T_{c}$ being the critical temperature value; and $b(=2c_{s}/\rho_{0})$ is a coupling constant related to the speed of sound $c_s$ and to the background fluid density $\rho_0$. An acoustic black hole metric can be obtained by considering perturbations (fluctuations of the complex scalar field) around the background in the Madelung representation $\varphi=\mbox{e}^{i\theta(\mathbf{x},t)}\sqrt{\rho(\mathbf{x},t)}$, which after some algebra (for details, refer to Ref.~\cite{PhysRevD.99.104047} and references therein) gives rise to a relativistic wave equation (similar to the massless Klein-Gordon equation) that governs the propagation of the phase fluctuations (weak excitations in a homogeneous stationary condensate).

Thus, the line element describing static dumb black holes, the so-called Schwarzschild acoustic black holes, can be written as
\begin{equation}
ds^{2}=g_{\mu\nu}dx^{\mu}dx^{\nu}=\sqrt{3}c_{s}^{2}\biggl[-f(r)\ dt^{2}+\frac{1}{f(r)}\ dr^{2}+r^{2}\ d\theta^{2}+r^{2}\sin^{2}\theta\ d\phi^{2}\biggr],
\label{eq:Schw_acous_metric}
\end{equation}
where the function $f(r)$ has the form
\begin{equation}
f(r)=\biggl(1-\frac{2M}{r}\biggr)\biggl[1-\xi\frac{2M}{r}\biggl(1-\frac{2M}{r}\biggr)\biggr].
\label{eq:f(r)_SABH}
\end{equation}
Here, $M$ is the total mass centered at the origin of the system of coordinates, and $\xi$ is the tuning parameter related to the radial component of the background fluid 4-velocity, $v_{r}=\sqrt{2M\xi/r}$.  For the velocity to be real, $\xi > 0$ (see Ref.~\cite{PhysRevD.99.104047} for a review, and references therein). It is obvious that at the limit $\xi \rightarrow 0$ the well-known Schwarzschild black hole is recovered, with an event horizon of radius $r_{\rm s}=2M$. On the other hand, $\xi \rightarrow \infty$ implies that the escape velocity $v_{r}$ goes to infinity and hence an acoustic black hole covers the whole spacetime. For this reason, $\xi > 0$ is defined as the tuning parameter, for this type of acoustic black holes.

The horizons of the Schwarzschild acoustic black hole (SABH) are given by the zeros of $f(r)=0$, that is,
\begin{equation}
f(r)=0=(r-r_{\rm s})(r-r_{\rm ac_{-}})(r-r_{\rm ac_{+}}).
\label{eq:surface_SABH}
\end{equation}
The solutions of this equation are as follows. The ``optical'' event horizon is $r_{\rm s}=2M$, while the ``interior'' and ``exterior'' acoustic event horizons are $r_{\rm ac_{-}}=M(\xi-\sqrt{\xi^{2}-4\xi})$ and $r_{\rm ac_{+}}=M(\xi+\sqrt{\xi^{2}-4\xi})$, respectively. Note that when $\xi=4$, we get an extreme Schwarzschild acoustic black hole; that is, the interior and exterior horizons coincide at $r_{\rm e}=4M$. This observation leads to the necessary condition for the existence of acoustic event horizons, that is $\xi\ge 4$. The exterior event horizon $r_{\rm ac_{+}}$ is the outermost marginally trapped surface for the outgoing phonons. Indeed, that is the last surface from which both light and sound waves could still escape from the acoustic black hole. Thus, it is meaningful to study quantum particles that propagate outside the exterior event horizon, whose equation of motion is discussed in the next section.
%
%%%%%%%%%%%%%%%%%%%%%%%%%%%%%%%%%%%%%%%%%%%%%%%%%%%%%%%%%%%%%%%%%%%%%%%%%%%%%%%%%%%%%%%%%%%%%% Klein-Gordon equation
%
\section{Klein-Gordon equation}\label{III}
It was also shown by Ge \textit{et al.} \cite{PhysRevD.99.104047} that the metric (\ref{eq:Schw_acous_metric}) describing SABHs obeys the covariant massless Klein-Gordon equation, which is given by
\begin{equation}
\biggl[\frac{1}{\sqrt{-g}}\partial_{\mu}(g^{\mu\nu}\sqrt{-g}\partial_{\nu})\biggl]\Psi=0,
\label{eq:massless_Klein-Gordon}
\end{equation}
where $\Psi=\Psi(t,r,\theta,\phi)$ is the scalar wave function, and $g \equiv \mbox{det}(g_{\mu\nu})$. Note that the scalar wave function will not depend on the speed of sound, since the term $\sqrt{3} \, c_{s}^{2}$ is a kind of ``conformal factor'' on the metric given by Eq.~(\ref{eq:Schw_acous_metric}), which means that we can choose $c_{s}^{2}=1/\sqrt{3}$ without loss of generality.

The equations of motion (\ref{eq:massless_Klein-Gordon}) for the metric (\ref{eq:Schw_acous_metric}) lead in a system of two separated ordinary differential equations for the angular and radial parts of the scalar wave function, which due to the spherical symmetry can be written as $\Psi(t,r,\theta,\phi)=\mbox{e}^{-i \omega t}R(r)Y_{lm}(\theta,\phi)$. They are given by
\begin{equation}
\frac{1}{\sin^{2}\theta}\frac{\partial^{2} Y_{lm}(\theta,\phi)}{\partial \phi^{2}}+\frac{1}{\sin\theta}\frac{\partial}{\partial \theta}\biggl[\sin\theta\frac{\partial Y_{lm}(\theta,\phi)}{\partial \theta}\biggr]=0
\label{eq:angular_SABH}
\end{equation}
and
\begin{equation}
\frac{d^{2} R(r)}{d r^{2}}+\frac{1}{F(r)}\frac{d F(r)}{d r}\frac{d R(r)}{d r}+\biggl\{\frac{r^{4}\omega^{2}}{[F(r)]^{2}}-\frac{\lambda}{F(r)}\biggr\}R(r)=0,
\label{eq:radial_SABH}
\end{equation}
where $\lambda=l(l+1)$ is a separation constant, $\omega$ is the frequency (energy) of the scalar particle, $Y_{lm}(\theta,\phi)$ are the spherical harmonic functions, $R(r)$ is the radial function, and $F(r)=r^{2}f(r)$.
%
%%%%%%%%%%%%%%%%%%%%%%%%%%%%%%%%%%%%%%%%%%%%%%%%%%%%%%%%%%%%%%%%%%%%%%%%%%%%%%%%%%%%%%%%%%%%%% Effective potential
%
\subsection{Effective potential}
Next, we would like to show the behavior of the effective potential, $V_{\rm eff}(r)$. By defining a new radial function $u(r)=R(r)/r$, Eq.~(\ref{eq:radial_SABH}) will be written as
\begin{equation}
\frac{d^{2} u(r)}{d r_{*}^{2}}+[\omega^{2}-V_{\rm eff}(r)]u(r)=0,
\label{eq:radial_u_SABH}
\end{equation}
with
\begin{equation}
V_{\rm eff}(r)=f(r)\biggl[\frac{\lambda}{r^{2}}+\frac{1}{r}\frac{df(r)}{dr}\biggr],
\label{eq:Veff_SABH}
\end{equation}
where we have introduced the ``tortoise coordinate'' $r_{*}$, which is defined by the relation $dr_{*}=dr/f(r)$.

\begin{figure}%[ht]
\centering
\includegraphics[width=0.66\columnwidth]{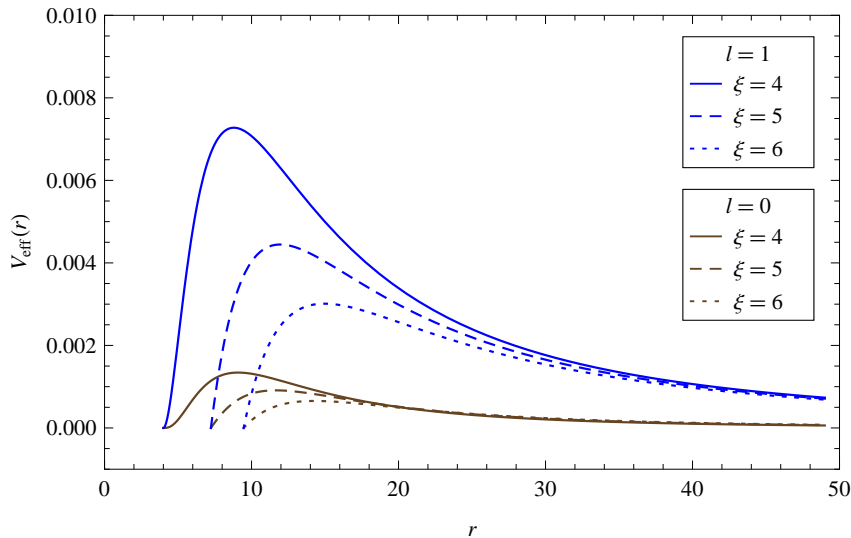}
\caption{The effective potentials for $M=1$, and different values of the azimuthal quantum number $l(=0,1)$ and the tuning parameter $\xi(=4,5,6)$.}
\label{fig:Fig1}
\end{figure}

Equation (\ref{eq:radial_u_SABH}) looks like a one-dimensional Schr\"{o}dinger equation, with an effective potential $V_{\rm eff}(r)$ given by Eq.~(\ref{eq:Veff_SABH}). The behavior of $V_{\rm eff}(r)$ is shown in Fig.~\ref{fig:Fig1}, for different values of the azimuthal quantum number and the tuning parameter.

From Eq.~(\ref{eq:Veff_SABH}) and Fig.~\ref{fig:Fig1}, we see that the width of the potential barrier increases with the azimuthal quantum number $l$, while decreasing with the tuning parameter $\xi$. For the chosen values, the exterior acoustic event horizon $r_{\rm ac_{+}}$ is at $r=4.0$, $r=7.2361$, and $r=9.4641$ for $\xi=4$, $\xi=5$, and $\xi=6$, respectively.
%
%%%%%%%%%%%%%%%%%%%%%%%%%%%%%%%%%%%%%%%%%%%%%%%%%%%%%%%%%%%%%%%%%%%%%%%%%%%%%%%%%%%%%%%%%%%%%% Radial equation
%
\subsection{Radial equation}
Here, we will provide an analytic, general solution for the radial part of the covariant massless Klein-Gordon equation in the SABH spacetime.

First, we use Eq.~(\ref{eq:surface_SABH}), which provides the values for the horizons, in order to transform the radial equation (\ref{eq:radial_SABH}) as
\begin{eqnarray}
&& \frac{d^{2} R}{d r^{2}}+\biggl(\frac{1}{r-r_{1}}+\frac{1}{r-r_{2}}+\frac{1}{r-r_{3}}\biggr)\frac{d R}{d r}+\biggl[\frac{r^4 \omega ^2-\lambda  (r-r_{1}) (r-r_{2}) (r-r_{3})}{(r-r_{1})^2 (r-r_{2})^2 (r-r_{3})^2}\biggr]R=0,
\label{eq:radial_2_SABH}
\end{eqnarray}
where, for convenience, we have renamed $r_{\rm ac_{+}}$, $r_{\rm ac_{-}}$, and $r_{\rm s}$ as $r_{1}$, $r_{2}$, and $r_{3}$, respectively.

The new form of the radial equation (\ref{eq:radial_2_SABH}) indicates the existence of three finite regular singularities associated with the three horizons $r_{1}$, $r_{2}$, and $r_{3}$ and a regular singularity at infinity, which implies that (\ref{eq:radial_2_SABH}) is a Heun-type equation. Therefore, we can transform this radial equation to a more suitable form, by defining a new radial coordinate $x$ related to $r_{1}$, $r_{2}$, and $r_{3}$ as
\begin{equation}
x=\frac{r-r_{1}}{r_{2}-r_{1}},
\label{eq:radial_coordinate_SABH}
\end{equation}
and we set a new parameter $a$ associated with the three finite regular singularities as
\begin{equation}
a=\frac{r_{3}-r_{1}}{r_{2}-r_{1}}.
\label{eq:a_SABH}
\end{equation}
Thus, by substituting Eqs.~(\ref{eq:radial_coordinate_SABH}) and (\ref{eq:a_SABH}) into Eq.~(\ref{eq:radial_2_SABH}), we obtain
\begin{equation}
\frac{d^{2} R(x)}{d x^{2}}+\biggl(\frac{1}{x}+\frac{1}{x-1}+\frac{1}{x-a}\biggr)\frac{d R(x)}{d x}+\biggl[\sum_{j=0}^{2}\frac{L_{j}}{x-x_{j}}+\sum_{j=0}^{2}\frac{Q_{j}}{(x-x_{j})^{2}}\biggr]R(x)=0,
\label{eq:radial_x_SABH}
\end{equation}
where $j=0,1,2$ labels the (now shifted) singularities $x_{j}=(0,1,a)$ and the coefficients $L_{j}$ and $Q_{j}$ are given by
\begin{equation}
L_{0}=\frac{2 r_{1}^3 \omega ^2 [r_{1} (r_{2}+r_{3})-2 r_{2} r_{3}]}{(r_{1}-r_{2})^2 (r_{1}-r_{3})^3}+\frac{\lambda }{r_{1}-r_{3}},
\label{eq:L0_SABH}
\end{equation}
\begin{equation}
L_{1}=-\frac{2 r_{2}^3 \omega ^2 [r_{1} (r_{2}-2 r_{3})+r_{2} r_{3}]}{(r_{1}-r_{2})^2 (r_{2}-r_{3})^3}-\frac{\lambda }{r_{2}-r_{3}},
\label{eq:L1_SABH}
\end{equation}
\begin{equation}
L_{2}=-\frac{(r_{1}-r_{2}) \{2 r_{3}^3 \omega ^2 [r_{3} (r_{1}+r_{2})-2 r_{1} r_{2}]+\lambda  (r_{1}-r_{3})^2 (r_{2}-r_{3})^2\}}{(r_{1}-r_{3})^3 (r_{3}-r_{2})^3},
\label{eq:L2_SABH}
\end{equation}
\begin{equation}
Q_{0}=\frac{r_{1}^4 \omega^2}{(r_{1}-r_{2})^2(r_{1}-r_{3})^2},
\label{eq:Q0_SABH}
\end{equation}
\begin{equation}
Q_{1}=\frac{r_{2}^4 \omega^2}{(r_{1}-r_{2})^2(r_{2}-r_{3})^2},
\label{eq:Q1_SABH}
\end{equation}
\begin{equation}
Q_{2}=\frac{r_{3}^4 \omega^2}{(r_{3}-r_{1})^2(r_{3}-r_{2})^2}.
\label{eq:Q2_SABH}
\end{equation}

Now, Eq.~(\ref{eq:radial_x_SABH}) is almost similar to the general Heun equation, where we just need to reduce the power of the terms $1/(x-x_{j})^{2}$. To do this, finally, we perform a \textit{F-homotopic transformation} of the dependent function, $R(x) \mapsto U(x)$, such that
\begin{equation}
R(x)=x^{A_{0}}(x-1)^{A_{1}}(x-a)^{A_{2}}U(x),
\label{eq:F-homotopic_SABH}
\end{equation}
where the coefficients $A_{j}$ are the exponents of the three singular points $x_{j}$ in Eq.~(\ref{eq:radial_x_SABH}). They obey to the following indicial equation:
\begin{equation}
G(z)=z(z-1)+z+Q_{j}=z^{2}+Q_{j}=0.
\label{eq:indicial_equation}
\end{equation}
Their roots are given by
\begin{equation}
z_{1,2}^{x=0}=\pm \frac{i r_{1}^2 \omega}{(r_{1}-r_{2})(r_{1}-r_{3})} \equiv A_{0},
\label{eq:A0_SABH}
\end{equation}
\begin{equation}
z_{1,2}^{x=1}=\pm \frac{i r_{2}^2 \omega}{(r_{1}-r_{2})(r_{2}-r_{3})} \equiv A_{1},
\label{eq:A1_SABH}
\end{equation}
\begin{equation}
z_{1,2}^{x=a}=\pm \frac{i r_{3}^2 \omega}{(r_{3}-r_{1})(r_{3}-r_{2})} \equiv A_{2}.
\label{eq:A2_SABH}
\end{equation}

Then, by substituting Eqs.~(\ref{eq:F-homotopic_SABH})--(\ref{eq:A2_SABH}) into Eq.~(\ref{eq:radial_x_SABH}), we derive a new equation for the radial function $U(x)$,
\begin{equation}
\frac{d^{2}U(x)}{dx^{2}}+\biggl(\frac{1+2A_{0}}{x}+\frac{1+2A_{1}}{x-1}+\frac{1+2A_{2}}{x-a}\biggr)\frac{dU(x)}{dx}+\frac{A_{3}x-A_{4}}{x(x-1)(x-a)}U(x)=0,
\label{eq:radial_3_SABH}
\end{equation}
with
\begin{equation}
A_{3}=-a (L_{0}+L_{1})+2 A_{0} (A_{1}+A_{2}+1)+2 A_{1} (A_{2}+1)+2 A_{2}-L_{0}-L_{2},
\label{eq:A3_SABH}
\end{equation}
\begin{equation}
A_{4}=A_{0} (2 a A_{1}+a+2 A_{2}+1)+a (A_{1}-L_{0})+A_{2}.
\label{eq:A4_SABH}
\end{equation}
By these substitutions, the three singularities $r_1$, $r_2$, and $r_3$ were moved to the points $x=0$, $1$, and $a$, respectively, and we can conclude that Eq.~(\ref{eq:radial_3_SABH}) has the form of a general Heun equation \cite{Ronveaux:1995}, whose canonical form is given by
\begin{equation}
\frac{d^{2}U(x)}{dx^{2}}+\left(\frac{\gamma}{x}+\frac{\delta}{x-1}+\frac{\epsilon}{x-a}\right)\frac{dU(x)}{dx}+\frac{\alpha\beta x-q}{x(x-1)(x-a)}U(x)=0.
\label{eq:general_Heun_canonical_form}
\end{equation}
Now the radial function $U(x)$ becomes the general Heun function $U(x) \equiv \mbox{HeunG}(a,q;\alpha,\beta,\gamma,\delta;x)$, which is simultaneously a Frobenius solution around two singularities $s_{1}$ and $s_{2}$, where $s_{1},s_{2} \in \{0,1,a\}$ and is analytic in some domain including both these singularities.

The parameters on which the Heun function depends are in general complex and play different roles. Namely, $a (\neq 0,1)$ is the \emph{singularity parameter}; $q$ is the \emph{accessory parameter}; and $\alpha$, $\beta$, $\gamma$, $\delta$, and $\epsilon$ are the \emph{exponent parameters}, which are related by $\gamma+\delta+\epsilon=\alpha+\beta+1$.

Therefore, an analytical, general solution for the radial part of the covariant massless Klein-Gordon equation, in the SABH spacetime, can be written as
\begin{eqnarray}
R(x) & = & x^{A_{0}}(x-1)^{A_{1}}(x-a)^{A_{2}}\nonumber\\
& & \times [C_{1}\ \mbox{HeunG}(a,q;\alpha,\beta,\gamma,\delta;x)+C_{2}\ x^{1-\gamma}\ \mbox{HeunG}(a,q_{2};\alpha_{2},\beta_{2},\gamma_{2},\delta;x)],
\label{eq:general_solution_radial_3_SABH}
\end{eqnarray}
where $C_{1}$ and $C_{2}$ are constants to be determined. These are two linearly independent solutions of the general Heun equation since $\gamma \neq 0,-1,-2,\ldots$ (in the first solution) and $\gamma \neq 1,2,\ldots$ (in the second solution), corresponding to the exponents $0$ and $1-\gamma$ at $x=0$. The parameters $\alpha$ and $\beta$ are the roots $y_{+}$ and $y_{-}$, respectively, of the equation $y^{2}+(1-\gamma-\delta-\epsilon)y+A_{3}=0$ \cite{Ronveaux:1995}, given by
\begin{equation}
\alpha=1+A_{0}+A_{1}+A_{2}+\sqrt{a(L_{0}+L_{1})+A_{0}^2+A_{1}^2+A_{2}^2+L_{0}+L_{2}+1},
\label{eq:alpha_SABH}
\end{equation}
\begin{equation}
\beta=1+A_{0}+A_{1}+A_{2}-\sqrt{a(L_{0}+L_{1})+A_{0}^2+A_{1}^2+A_{2}^2+L_{0}+L_{2}+1}.
\label{eq:beta_SABH}
\end{equation}
The other parameters, namely, $\gamma$, $\delta$, $\epsilon$, and $q$ are given by
\begin{equation}
\gamma=1+2A_{0},
\label{eq:gamma_SABH}
\end{equation}
\begin{equation}
\delta=1+2A_{1},
\label{eq:delta_SABH}
\end{equation}
\begin{equation}
\epsilon=1+2A_{2},
\label{eq:eta_SABH}
\end{equation}
\begin{equation}
q=A_{4}.
\label{eq:q_SABH}
\end{equation}
Furthermore, for the second solution, the parameters $\alpha_{2}$, $\beta_{2}$, $\gamma_{2}$, and $q_{2}$ are given by
\begin{equation}
\alpha_{2}=\alpha+1-\gamma,
\label{eq:alpha_1_general_Heun}
\end{equation}
\begin{equation}
\beta_{2}=\beta+1-\gamma,
\label{eq:beta_1_general_Heun}
\end{equation}
\begin{equation}
\gamma_{2}=2-\gamma,
\label{eq:gamma_1_general_Heun}
\end{equation}
\begin{equation}
q_{2}=q+(\alpha\delta+\epsilon)(1-\gamma).
\label{eq:q_1_general_Heun}
\end{equation}
Note that the final expressions for these parameters will depend on the signs to be chosen for the coefficients $A_{j}$ given by Eqs.~(\ref{eq:A0_SABH})--(\ref{eq:A2_SABH}). In this work, the negative sign is the correct choice, which will be justified, confirmed in the discussion of the quasibound states.

In what follows, we will use this analytical solution of the radial equation, in the SABH spacetime, and the properties of the general Heun functions to discuss the Hawking radiation, the quasibound state frequencies and its corresponding wave functions.
%
%%%%%%%%%%%%%%%%%%%%%%%%%%%%%%%%%%%%%%%%%%%%%%%%%%%%%%%%%%%%%%%%%%%%%%%%%%%%%%%%%%%%%%%%%%%%%% Hawking radiation
%
\section{Hawking radiation}\label{IV}
In this section, we will obtain the Hawking radiation spectrum for massless scalar particles near the exterior event horizon of a SABH spacetime.

In the limit $r \rightarrow r_{1}$, which leads to $x \rightarrow 0$, we can calculate the corresponding Heun function. Thus, for $\gamma \neq 0,-1,-2,\ldots$, which corresponds to the first solution of our case, the general Heun functions are analytic in the disk $|x| < 1$, and the following Maclaurin expansion applies \cite{MathAnn.33.161},
\begin{equation}
\mbox{HeunG}(a,q;\alpha,\beta,\gamma,\delta;x)=\sum_{\nu=0}^{\infty}c_{\nu}x^{\nu},
\label{eq:serie_HeunG_todo_x}
\end{equation}
where
\begin{eqnarray}
-qc_{0}+a \gamma c_{1}&=&0,\nonumber\\
P_{\nu}c_{\nu-1}-(Q_{\nu}+q)c_{\nu}+X_{\nu}c_{\nu+1}&=&0 \quad (\nu \geq 1),
\label{eq:recursion_General_Heun}
\end{eqnarray}
with $c_{0}=1$ and
\begin{eqnarray}
P_{\nu}&=&(\nu-1+\alpha)(\nu-1+\beta),\nonumber\\
Q_{\nu}&=&\nu[(\nu-1+\gamma)(1+a)+a\delta+\epsilon],\nonumber\\
X_{\nu}&=&(\nu+1)(\nu+\gamma)a.
\label{eq:P_Q_X_recursion_General_Heun}
\end{eqnarray}
This expansion leads to $\mbox{HeunG}(a,q;\alpha,\beta,\gamma,\delta;0) \sim 1$. Similarly, for $\gamma \neq 1,2,\ldots$, it is easy to verify \cite{MathAnn.33.161} that we get the same behavior for the second solution. In this limit, the radial solution (\ref{eq:general_solution_radial_3_SABH}) has the following asymptotic behavior at the exterior event horizon,
\begin{equation}
R(r) \sim C_{1}\ (r-r_{1})^{A_{0}}+C_{2}\ (r-r_{1})^{-A_{0}},
\label{eq:asymptotic_SABH}
\end{equation}
where all remaining constants have been included in $C_{1}$ and $C_{2}$. Note that we recovered the original radial coordinate $r$. Thus, by including the time dependence, the solution can be written as
\begin{equation}
\Psi(r,t) \sim C_{1}\ \Psi_{in} + C_{2}\ \Psi_{out},
\label{eq:full_wave_SABH}
\end{equation}
where the ingoing and outgoing scalar wave solutions are given, respectively, by
\begin{equation}
\Psi_{in}(r>r_{1})=\mbox{e}^{-i \omega t}(r-r_{1})^{-\frac{i\omega}{2\kappa_{1}}}
\label{eq:sol_in_1_SABH}
\end{equation}
and
\begin{equation}
\Psi_{out}(r>r_{1})=\mbox{e}^{-i \omega t}(r-r_{1})^{+\frac{i\omega}{2\kappa_{1}}}.
\label{eq:sol_out_2_SABH}
\end{equation}
Here, $\kappa_{1}$ is the gravitational acceleration on the exterior horizon $r_{1}$ and is given by
\begin{equation}
\kappa_{1} \equiv \frac{1}{2r_{1}^{2}} \left.\frac{dF(r)}{dr}\right|_{r=r_{1}} = \frac{(r_{1}-r_{2})(r_{1}-r_{3})}{2r_{1}^{2}}
\label{eq:grav_acc_SABH}
\end{equation}
such that, from Eq.~(\ref{eq:A0_SABH}) with negative sign, we get
\begin{equation}
A_{0}=-\frac{i\omega}{2\kappa_{1}}.
\label{eq:A0_full_wave_SABH}
\end{equation}

Finally, by following Vieira et \textit{al.} \cite{AnnPhys.350.14}, we can obtain the relative scattering probability, $\Gamma_{1}$, and the Hawking radiation spectra, $\bar{N}_{\omega}$, which are given, respectively, by
\begin{equation}
\Gamma_{1}=\left|\frac{\Psi_{\rm out}(r>r_{1})}{\Psi_{\rm out}(r<r_{1})}\right|^{2}=\mbox{e}^{-\frac{2\pi\omega}{\kappa_{1}}}
\label{eq:rel_prob_SABH}
\end{equation}
and
\begin{equation}
\bar{N}_{\omega}=\frac{\Gamma_{1}}{1-\Gamma_{1}}=\frac{1}{\mbox{e}^{\hbar\omega/k_{B}T_{1}}-1}.
\label{eq:rad_spec_SABH}
\end{equation}
This means that the resulting radiation spectrum, for massless scalar particles in the background under consideration, has a thermal character, which is analogous to the blackbody spectrum. Note that here we have used the definition of the Hawking-Unruh temperature, namely, $k_{B}T_{1}=\hbar\kappa_{1}/2\pi$, where $k_{B}$ is the Boltzmann constant.
%
%%%%%%%%%%%%%%%%%%%%%%%%%%%%%%%%%%%%%%%%%%%%%%%%%%%%%%%%%%%%%%%%%%%%%%%%%%%%%%%%%%%%%%%%%%%%%% Quasibound states
%
\section{Quasibound states}\label{V}
In order to investigate the quasibound states, which are solutions to the equation of motion that tend to zero at infinity, we will calculate the spectrum of resonant frequencies on the background under consideration.

The quasibound states, also referred to in the literature as resonance spectra or quasistationary levels, are localized in the black hole potential well, and tend to zero at spatial infinity. This means that there exist two boundary conditions associated to the spectrum of quasibound states. Since the flux of particles crosses into the exterior horizon surface, the spectrum of quasibound states has complex frequencies, so that it can be expressed as $\omega_{n}=\omega_{R}+i\omega_{I}$, where $\omega_{R}=\mbox{Re}[\omega_{n}]$ and $\omega_{I}=\mbox{Im}[\omega_{n}]$ are the real and imaginary parts, respectively, and $n$ is the overtone number. The sign of the imaginary part shows whether the wave solution decays ($\mbox{Im}[\omega_{n}] < 0$) or grows ($\mbox{Im}[\omega_{n}] > 0$) with the time.

Then, it is possible, in  principle, to derive the characteristic resonance equation by solving the radial equation in two different asymptotic regions and using a standard matching procedure for these two radial solutions in their common overlap region. These ideas have been explored by many authors in the last years \cite{PhysRevD.85.044031,PhysLettB.749.167,PhysRevD.103.044062}. Here, we will obtain the spectrum of quasibound states by extending the method developed by Vieira and Bezerra \cite{AnnPhys.373.28} to find the resonant frequencies and then impose two boundary conditions to the radial solution.

First, we demand that the radial solution should describe an ingoing wave at the exterior event horizon. In order to fully satisfy this boundary condition, we must impose that $C_{2}=0$ in Eq.~(\ref{eq:full_wave_SABH}), and in (\ref{eq:general_solution_radial_3_SABH}) as well.

Second, it is necessary that the radial solution should tend to zero far from the black hole at asymptotic infinity. In this limit, we will use the two linearly independent solutions of the general Heun equation at $x=\infty$, which correspond to the exponents $\alpha$ and $\beta$, to write the radial solution (\ref{eq:general_solution_radial_3_SABH}) as
\begin{eqnarray}
R(x) & = & x^{A_{0}}(x-1)^{A_{1}}(x-a)^{A_{2}}\nonumber\\
& & \times \biggl[C_{1}\ x^{-\alpha}\ \mbox{HeunG}\biggl(\frac{1}{a},\alpha(\beta-\epsilon)+\frac{\alpha}{a}(\beta-\delta)-\frac{q}{a};\alpha,\alpha-\gamma+1,\alpha-\beta+1,\delta;\frac{1}{x}\biggr)\nonumber\\
&& + C_{2}\ x^{-\beta}\ \mbox{HeunG}\biggl(\frac{1}{a},\beta(\alpha-\epsilon)+\frac{\beta}{a}(\alpha-\delta)-\frac{q}{a};\beta,\beta-\gamma+1,\beta-\alpha+1,\delta;\frac{1}{x}\biggr)\biggr],
\label{eq:radial_solution_infinity_SABH}
\end{eqnarray}
Thus, in the limit when $r \rightarrow \infty$, which implies that $x \rightarrow \infty$, by choosing the negative sign in Eqs.~(\ref{eq:A0_SABH})--(\ref{eq:A2_SABH}), the radial solution (\ref{eq:radial_solution_infinity_SABH}) has the following asymptotic behavior,
\begin{equation}
R(r) \sim C_{1}\ \frac{1}{r^{i B \omega+\alpha}}+C_{2}\ \frac{1}{r^{i B \omega+\beta}}
\label{eq:radial_infinity_SABH}
\end{equation}
where the coefficient $B$ is given by
\begin{equation}
B =1+\frac{2\xi}{\sqrt{\xi(\xi-4)}}.
\label{eq:B_SABH}
\end{equation}
However, since $C_{2}=0$, we get
\begin{equation}
R(r) \sim C_{1}\ \frac{1}{r^{\sigma}}.
\label{eq:radial_infinity_2_SABH}
\end{equation}
where $\sigma=i B \omega+\alpha$. Note that, since $\xi > 4$ (nonextreme case), the coefficient $B$ is such that $B > 1$. The sign of the real part of $\sigma$ determines the behavior of the wave function as $r \rightarrow \infty$. If $\mbox{Re}[\sigma] > 0$, the solution tends to zero, whereas if $\mbox{Re}[\sigma] < 0$, the solution diverges; by definition, the quasibound state solutions are ingoing at the horizon, and tend to zero at infinity ($\mbox{Re}[\sigma] > 0$). The final behavior of the scalar wave function will be determined when we know the values of the frequency $\omega$ and the parameter $\alpha$, which will be obtained in the next.
%
%%%%%%%%%%%%%%%%%%%%%%%%%%%%%%%%%%%%%%%%%%%%%%%%%%%%%%%%%%%%%%%%%%%%%%%%%%%%%%%%%%%%%%%%%%%%%% Polynomial condition
%
\subsection{Polynomial condition}
Now, we will show that the quasibound states of massless scalar fields in the SABHs can be found by imposing a polynomial condition, which gives rise to the general Heun polynomials.

It is known that a Heun polynomial is the solution of the Heun's differential equations which is valid at three singularities, in the sense of being simultaneously a Frobenius solution at each one of them \cite{Ronveaux:1995}. Therefore, in order to satisfy the second boundary condition, it is necessary that the radial solution must be written in terms of the Heun polynomials, which means that we need to derive a form of the Heun's functions that present a polynomial behavior; it will be done in the next section. In fact, that is a kind of standard matching procedure.

In this sense, we will use the fact that the general Heun functions become polynomials of degree $n$ if they satisfy the so-called $\alpha$-condition \cite{Ronveaux:1995}, which is given by
\begin{equation}
\alpha=-n,
\label{eq:alpha-condition}
\end{equation}
where $n=0,1,2,\ldots$ is now the principal quantum number. Such polynomial solutions are denoted by $\mbox{Hp}_{n}(x)=\mbox{HeunG}(a,q;-n,\beta,\gamma,\delta;x)$, and their properties will be discussed in the next section.

In our case, from Eqs.~(\ref{eq:A0_SABH})--(\ref{eq:A2_SABH}) (with the negative sign) and (\ref{eq:alpha_SABH}), the parameter $\alpha$ can be simply written as
\begin{equation}
\alpha=1-iB \omega+\sqrt{1-\omega^{2}}.
\label{eq:alpha_2_SABH}
\end{equation}
Then, by imposing the polynomial condition given by Eq.~(\ref{eq:alpha-condition}), we obtain the following expressions for the massless scalar resonant frequencies:
\begin{equation}
\omega_{n}^{\ (\pm)} = -i\frac{(n+1)B \pm \sqrt{n(n+2)+B^{2}}}{B^{2}-1}.
\label{eq:omega_SABH}
\end{equation}
In fact, these eigenvalues are the solutions of a second order equation for $\omega$; from now on, we will refer to them as the ``plus'' and ``minus'' solution. Note that the obtained quasispectrum is purely imaginary and always negative, as the tuning parameter $\xi$ and the principal quantum number $n$ grow. This corresponds to a trajectory that decays without oscillating and therefore we are dealing with an overdamped motion.

Some characteristic values of the coefficients $B$ and $\sigma$, as well as the corresponding massless scalar resonant frequencies, are shown in Table \ref{tab:I_SABH}, as functions of the tuning parameter $\xi$.

\begin{table}%[hb]
\caption{Values of the coefficients $B$ and $\sigma_{n}^{\ (\pm)}=i B \omega_{n}^{\ (\pm)}-n$, and the corresponding resonant frequencies $\omega_{n}^{\ (\pm)}$.}
\label{tab:I_SABH}
\begin{tabular}{c||c||c|c||c|c||c|c||c|c}
\hline\noalign{\smallskip}
			$\xi$    & $B$        & $\omega_{0}^{\ (+)}$ & $\sigma_{0}^{\ (+)}$ & $\omega_{1}^{\ (+)}$ & $\sigma_{1}^{\ (+)}$ & $\omega_{0}^{\ (-)}$        & $\sigma_{0}^{\ (-)}$ & $\omega_{1}^{\ (-)}$ & $\sigma_{1}^{\ (-)}$ \\
\noalign{\smallskip}\hline\noalign{\smallskip}
			4.01     & $41.05000$ & $-0.048750i$         & $2.001190$         & $-0.073146i$           & $2.002670$           & $0$             & $0$                  & $-0.024353i$         & $-0.000296$ \\
			5        & $5.472140$ & $-0.378115i$         & $2.069100$         & $-0.576417i$           & $2.154230$           & $0$             & $0$                  & $-0.179813i$         & $-0.016038$ \\
			6        & $4.464100$ & $-0.471688i$         & $2.105660$         & $-0.724662i$           & $2.234960$           & $0$             & $0$                  & $-0.218714i$         & $-0.023639$ \\
			7        & $4.055050$ & $-0.525149i$         & $2.129500$         & $-0.810673i$           & $2.287320$           & $0$             & $0$                  & $-0.239625i$         & $-0.028309$ \\
			8        & $3.828430$ & $-0.560660i$         & $2.146450$         & $-0.868345i$           & $2.324400$           & $0$             & $0$                  & $-0.252975i$         & $-0.031502$ \\
			9        & $3.683280$ & $-0.586203i$         & $2.159150$         & $-0.910095i$           & $2.352140$           & $0$             & $0$                  & $-0.262312i$         & $-0.033831$ \\
			10       & $3.581990$ & $-0.605544i$         & $2.169050$         & $-0.941855i$           & $2.373710$           & $0$             & $0$                  & $-0.269233i$         & $-0.035609$ \\
			$\vdots$ & $\vdots$   & $\vdots$             & $\vdots$             & $\vdots$             & $\vdots$             & $\vdots$             & $\vdots$             & $\vdots$             & $\vdots$   \\
			$\infty$ & $3.000000$ & $-0.750000i$         & $2.250000$         & $-1.183010i$           & $2.549040$           & $0$             & $0$                  & $-0.316987i$         & $-0.049038$ \\
\noalign{\smallskip}\hline
\end{tabular}
\end{table}

In Table \ref{tab:I_SABH}, we cannot determine the value of the coefficient $B$ nor the resonant frequencies, for the extreme case when $\xi=4$, since there exists a discontinuity in that point, which is due to the fact that, when we set $r_{1}=r_{2}$, the singularity parameter $a$, given by Eq.~(\ref{eq:a_SABH}), goes to infinity and hence the radial solution must be recalculated in this new scenario; in this limit, the coefficient $A_{3}$, which corresponds to the term $\alpha\beta$ in the general Heun equation, goes to zero.

From Table \ref{tab:I_SABH}, we conclude that the minus solution for the massless scalar resonant frequencies, $\omega_{n}^{\ (-)}$, is not a physically acceptable solution, since $\mbox{Re}[\sigma_{n}^{\ (-)}] \leq 0$ and thus it does not allow solutions tending to zero at infinity. Therefore, the unique physically admissible resonant frequencies are the plus solution, $\omega_{n}^{\ (+)}$, which represent the quasibound state energies for massless scalar particles in the SABH spacetime. In this case, the radial solution (\ref{eq:general_solution_radial_3_SABH}) tends to zero far from the SABHs at asymptotic infinity, since $\mbox{Re}[\sigma_{n}^{\ (+)}] > 0$, as required by the conditions for quasibound states.

\begin{figure}%[ht]
\centering
\includegraphics[width=0.66\columnwidth]{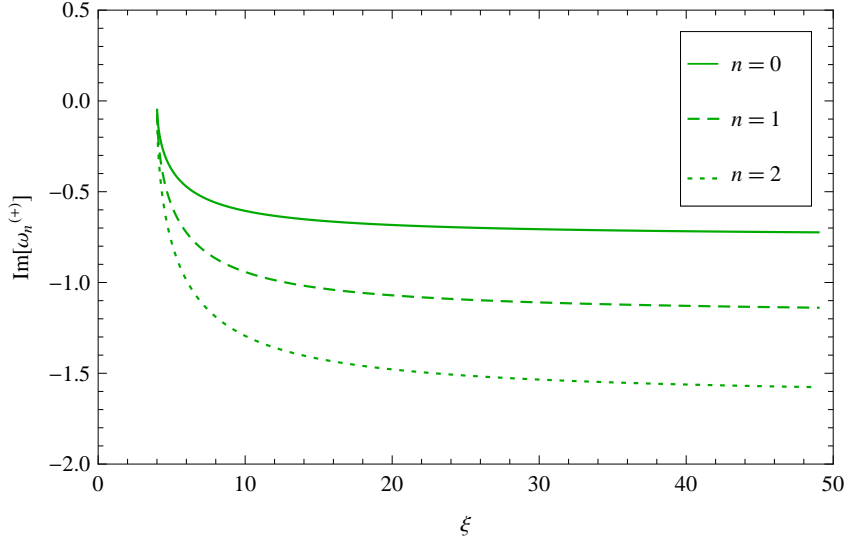}
\caption{The imaginary part of the massless scalar resonant frequencies $\omega_{n}^{\ (+)}$.}
\label{fig:Fig2}
\end{figure}

The behavior of the massless scalar resonant frequencies $\omega_{n}^{\ (+)}$ is shown in Fig.~\ref{fig:Fig2}, as functions of the tuning parameter $\xi$.

In Fig.~\ref{fig:Fig2}, we see that the imaginary part of the massless scalar resonant frequencies increases with the tuning parameter $\xi$, for fixed values of the principal quantum number $n$, but it reaches a limiting value when $\xi \rightarrow \infty$, as displayed in Table \ref{tab:I_SABH}.

In what concerns to the stability of the system, from Table \ref{tab:I_SABH} and Fig.~\ref{fig:Fig2}, we may conclude that the SABHs are stable, since the imaginary part of the massless scalar resonant frequencies is always negative.

It is worth pointing out that the resonant frequencies were obtained directly from the general form of Heun functions by using a polynomial condition, and, to our knowledge, there is no similar result in the literature for the background under consideration.
%
%%%%%%%%%%%%%%%%%%%%%%%%%%%%%%%%%%%%%%%%%%%%%%%%%%%%%%%%%%%%%%%%%%%%%%%%%%%%%%%%%%%%%%%%%%%%%% Schwarzschild black hole
%
\subsection{Schwarzschild black hole}
Now, we can compare these results with the ones known in the literature for the particular case of a standard Schwarzschild black hole.

In fact, the standard Schwarzschild black hole spacetime is recovered when $\xi=0$ and $c_{s}^{2}=1/\sqrt{3}$. Thus, it is possible, in principle, to obtain the massless scalar resonant frequencies for such a background by taking the limit $\xi \rightarrow 0$. However, that is not true. It is complicated, from the mathematical point of view, to manage the solution given in terms of the general Heun functions when some of their parameters go to zero (or infinity); a confluence process may be involved and hence the parameters of the new solution will be slightly different from the ones obtained by considering the original background from the beginning. Furthermore, from Eq.~(\ref{eq:B_SABH}), it is clear that the coefficient $B$ as well as the obtained resonant frequencies are not determined in the limit $\xi \rightarrow 0$.

Therefore, in order to compare our results with the ones concerning the Schwarzschild black hole, we will use the analytical solution of the Klein-Gordon equation and the expression for the field energies that was obtained by Muniz \textit{at al.} \cite{JCAP.01.006}, which concerns to the standard Schwarzschild black hole.

Thus, in a standard Schwarzschild black hole, the quasibound state frequencies of massless scalar particles are given by \cite{JCAP.01.006}
\begin{equation}
\omega_{n}=-\frac{i}{4M}(n+1),
\label{eq:Rfs_Schw}
\end{equation}
where $n=0,1,2,\ldots$. This quasispectrum is also purely imaginary and depends on the total mass $M$; it scales merely with mass, while for the SABHs we do not have such scaling.

We may compare these results as follows. Let us consider, for example, the fundamental mode $n=0$. An emitted signal, related to the massless scalar resonant frequencies, from a standard Schwarzschild black hole will be weaker (in intensity) than the one emitted by a SABH (for any value of the tuning parameter in this mode), that is, $\omega_{0} < \omega_{0}^{\ (+)} \ \forall \ \xi \geq 5$.
%
%%%%%%%%%%%%%%%%%%%%%%%%%%%%%%%%%%%%%%%%%%%%%%%%%%%%%%%%%%%%%%%%%%%%%%%%%%%%%%%%%%%%%%%%%%%%%% Wave functions
%
\section{Wave functions}\label{VI}
In this section, we will derive the eigenfunctions related to massless scalar particles propagating in the SABH background. This is possible if one uses some properties of the general Heun functions and then obtains their polynomial expressions.

The polynomial solutions of the general Heun equation (\ref{eq:general_Heun_canonical_form}) are denoted by $\mbox{Hp}_{n}(x)$ and can be written as
\begin{equation}
\mbox{Hp}_{n}(x)=\sum_{\nu=0}^{\infty}c_{\nu}x^{\nu},
\label{eq:polynomial_solutions}
\end{equation}
where the coefficients $c_{\nu}$ are given by the following set of equations,
\begin{eqnarray}
S_{0}c_{0}+X_{0}c_{1}&=&0,\nonumber\\
P_{\nu}c_{\nu-1}+S_{\nu}c_{\nu}+X_{\nu}c_{\nu+1}&=&0 \quad (\nu=1,2,\ldots,n-1),\nonumber\\
P_{n}c_{n-1}+S_{n}c_{n}&=&0,
\label{eq:cnu}
\end{eqnarray}
with $S_{\nu}=-Q_{\nu}-q$, and the expressions for $P_{\nu}$, $Q_{\nu}$, and $X_{\nu}$ are as given in (\ref{eq:P_Q_X_recursion_General_Heun}). These equations are consistent if and only if the accessory parameter $q$ was chosen properly, which means that it is calculated via a polynomial equation of degree $n+1$, namely, $c_{n+1}=0$. We will use for these eigenvalues the notation $q_{n,m}$, where $m$ runs from $0$ to $n$. Then, the corresponding general Heun polynomials will be denoted as $\mbox{Hp}_{n,m}(x)$.

In our case, the explicit form of the first two general Heun polynomials is obtained as follows. For $n=0$, we have
\begin{equation}
\mbox{Hp}_{0,m}(x)=c_{0}=1.
\label{eq:Hp0m}
\end{equation}
The eigenvalues $q_{0,m}$ must obey
\begin{equation}
c_{1}=0,
\end{equation}
where
\begin{equation}
-qc_{0}+a \gamma c_{1}=0,
\label{eq:q}
\end{equation}
which implies
\begin{equation}
c_{1}=\frac{q}{a \gamma},
\label{eq:c1}
\end{equation}
and then we have that
\begin{equation}
q_{0,0}=0.
\label{eq:q00}
\end{equation}
Thus, the general Heun polynomial for the fundamental mode is given by
\begin{equation}
\mbox{Hp}_{0,0}(x)=1.
\label{eq:Hp00}
\end{equation}
Now, for $n=1$, we have
\begin{equation}
\mbox{Hp}_{1,m}(x)=c_{0}+c_{1}x=1+\frac{q_{1,m}}{a \gamma}x.
\label{eq:Hp1m}
\end{equation}
Next, the eigenvalues $q_{1,m}$ must obey
\begin{equation}
c_{2}=0,
\end{equation}
where
\begin{equation}
P_{1}c_{0}-(Q_{1}+q)c_{1}+X_{1}c_{2}=0
\label{eq:P1}
\end{equation}
implying that
\begin{equation}
c_{2}=\frac{[\gamma(1+a)+a\delta+\epsilon+q]q-a \alpha\beta\gamma}{2a^{2}\gamma(1+\gamma)}
\label{eq:c2}
\end{equation}
and then we have that
\begin{equation}
q_{1,m}=\frac{-[\gamma(1+a)+a\delta+\epsilon] \pm \sqrt{\Delta}}{2},
\label{eq:q1m}
\end{equation}
with $\Delta=[\gamma(1+a)+a\delta+\epsilon]^{2}+4a \alpha\beta\gamma$. Here, the signs $-,+$ stand for $m=0,1$. Thus, the general Heun polynomials for the first excited mode are given by
\begin{equation}
\mbox{Hp}_{1,0}(x)=1+\frac{-[\gamma(1+a)+a\delta+\epsilon] - \sqrt{\Delta}}{2 a \gamma}x,
\label{eq:Hp10}
\end{equation}
\begin{equation}
\mbox{Hp}_{1,1}(x)=1+\frac{-[\gamma(1+a)+a\delta+\epsilon] + \sqrt{\Delta}}{2 a \gamma}x.
\label{eq:Hp11}
\end{equation}

Then, the radial eigenfunctions, for massless scalar particles propagating in a SABH spacetime, can be written as
\begin{equation}
R_{n,m}(x)=C_{n,m}\ x^{A_{0}}(x-1)^{A_{1}}(x-a)^{A_{2}}\ \mbox{Hp}_{n,m}(x),
\label{eq:eigenfunctions_SABH}
\end{equation}
where $C_{n,m}$ is a constant to be determined by using some additional boundary condition, as for example, that the wave function should be appropriately normalized in the range between the exterior event horizon and the infinity. It is worth emphasizing that these eigenfunctions are degenerate, due to the fact that the accessory parameter $q_{n,m}$ must be properly determined for each value of $n$.

\begin{figure}%[ht]
\centering
\includegraphics[width=0.66\columnwidth]{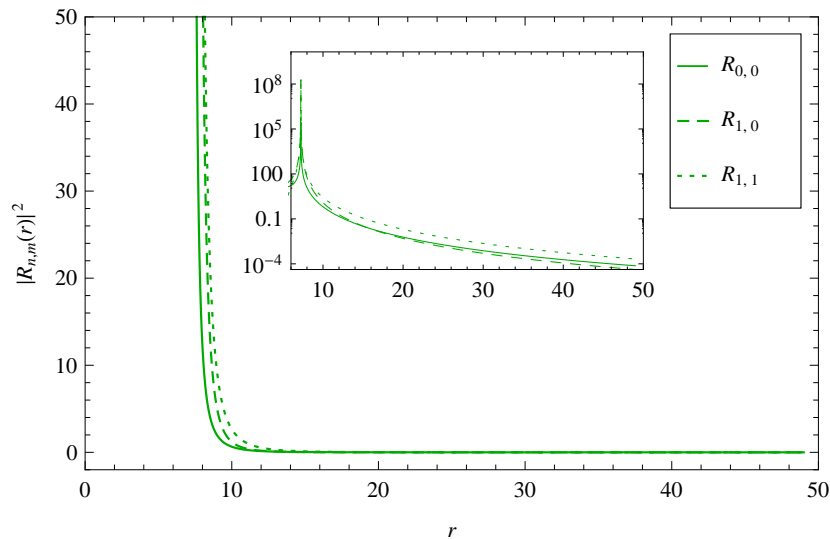}
\caption{The first three squared eigenfunctions related to $\omega_{n}^{\ (+)}$, for $M=1$. The units are in multiples of $C_{n,m}$.}
\label{fig:Fig3}
\end{figure}

Therefore, by using Eqs.~(\ref{eq:Hp00}), (\ref{eq:Hp10}), and (\ref{eq:Hp11}), we can plot the first three squared wave functions, which are presented in Fig.~\ref{fig:Fig3}.

From Fig.~\ref{fig:Fig3}, we conclude that the massless scalar resonant frequencies $\omega_{n}^{\ (+)}$ are quasibound states, since their wave functions present the desired behavior, that is, the radial solution tends to zero at infinity and diverges at the exterior event horizon; it (mathematically reaches a maximum value and then) crosses into the black hole.
%
%%%%%%%%%%%%%%%%%%%%%%%%%%%%%%%%%%%%%%%%%%%%%%%%%%%%%%%%%%%%%%%%%%%%%%%%%%%%%%%%%%%%%%%%%%%%%% Conclusions
%
\section{Conclusions}\label{VII}
In this work, we obtained an analytical, general solution for the covariant massless Klein-Gordon equation in the Schwarzschild acoustic black hole spacetime. The angular part of the solution is given in terms of the spherical harmonic functions, while the radial part of the solution is given in terms of the general Heun functions. The study of the radial solution led to some interesting physics.

We got the Hawking radiation spectrum for massless scalar particles in the vicinity to the exterior event horizon. We showed that this spectrum resembles the one describing a blackbody. It is worth emphasizing that the thermal Hawking-Unruh radiation was recently observed in an analog black hole \cite{NaturePhys.10.864,PhysRevD.92.024043,Nature.569.688}. Therefore, we present some analytical results that could be compared with detected data in a near future.

We obtained the quasispectrum of resonant frequencies for massless scalar particles propagating in the Schwarzschild acoustic black hole spacetime. This became possible by imposing two boundary conditions, and then we used a polynomial condition for the general Heun functions. We examined the behavior of the (polynomial) wave functions and showed that the massless scalar resonant frequencies $\omega_{n}^{\ (+)}$ describe quasistationary levels, and therefore they are quasibound states. We may conclude that the Schwarzschild acoustic black hole is stable, since the imaginary part of the massless scalar resonant frequencies $\omega_{n}^{\ (+)}$ is always negative.

Finally, the wave phenomena studied in this work are due to the interaction between quantum fields, in particular the scalar one, and the effective geometry of acoustic black holes in the Schwarzschild spacetime. Therefore, they are interesting semiclassical phenomena, which can give us some insights in the physics of black holes and for this reason should be investigated.
%
%%%%%%%%%%%%%%%%%%%%%%%%%%%%%%%%%%%%%%%%%%%%%%%%%%%%%%%%%%%%%%%%%%%%%%%%%%%%%%%%%%%%%%%%%%%%%% Author contributions
%
%\section*{Author contributions}
%H. S. Vieira: Conceptualization, investigation, methodology, software, writing.
%
%K. D. Kokkotas: Project administration, supervision, validation.
%
%%%%%%%%%%%%%%%%%%%%%%%%%%%%%%%%%%%%%%%%%%%%%%%%%%%%%%%%%%%%%%%%%%%%%%%%%%%%%%%%%%%%%%%%%%%%%% Data availability
%
\section*{Data availability}
The data that support the findings of this study are available from the corresponding author upon reasonable request.
%
%%%%%%%%%%%%%%%%%%%%%%%%%%%%%%%%%%%%%%%%%%%%%%%%%%%%%%%%%%%%%%%%%%%%%%%%%%%%%%%%%%%%%%%%%%%%%% Acknowledgments
%
\begin{acknowledgments}
H.S.V. is funded by the Alexander von Humboldt-Stiftung/Foundation (Grant No. 1209836). This study was financed in part by the Coordena\c c\~{a}o de Aperfei\c coamento de Pessoal de N\'{i}vel Superior - Brasil (CAPES) - Finance Code 001.
\end{acknowledgments}
%
%%%%%%%%%%%%%%%%%%%%%%%%%%%%%%%%%%%%%%%%%%%%%%%%%%%%%%%%%%%%%%%%%%%%%%%%%%%%%%%%%%%%%%%%%%%%%% thebibliography
%

%
%%%%%%%%%%%%%%%%%%%%%%%%%%%%%%%%%%%%%%%%%%%%%%%%%%%%%%%%%%%%%%%%%%%%%%%%%%%%%%%%%%%%%%%%%%%%%%
%
\end{document}